\documentclass[prl,twocolumn]{revtex4}
\usepackage{amssymb}
\usepackage{color}

\usepackage{amsthm}
\usepackage{graphicx}
\usepackage{dcolumn}
\usepackage{bm}
\usepackage{subfigure}
\usepackage{amssymb}
\usepackage{verbatim}
\usepackage{amscd}
\usepackage{amssymb}
\usepackage{amsmath, amsfonts}
\usepackage{setspace}
\usepackage[]{graphicx}        
\usepackage{amsthm}
\usepackage{enumerate}

\newcommand{\ba}[1]{\begin{align*} #1 \end{align*}}

\theoremstyle{plain}

\theoremstyle{plain}

\theoremstyle{plain}
 
\theoremstyle{plain}

\theoremstyle{remark}

\theoremstyle{conjecture}

\theoremstyle{observation}

\theoremstyle{definition}

\theoremstyle{corollary}

\theoremstyle{definition}

\theoremstyle{definition}

\theoremstyle{assumption}

\theoremstyle{definition}

\theoremstyle{problem}

\theoremstyle{fact}

\begin{document}

\title{Quantum criticality from Ising model on fractal lattices}
\author{Beni Yoshida}
\author{Aleksander Kubica}
\affiliation{Institute for Quantum Information and Matter, California Institute of Technology, Pasadena, California 91125, USA
}

\date{\today}
\begin{abstract}
We study the quantum Ising model on the Sierpi\'{n}ski triangle, whose Hausdorff dimension is $\log 3/ \log 2 \approx 1.585$, and demonstrate that it undergoes second-order phase transition with scaling relations satisfied precisely. We also study the quantum $3$-state Potts model on the Sierpi\'{n}ski triangle and find first-order phase transition, which is consistent with a prediction from $\epsilon$-expansion that the transition becomes first-order for $D > 1.3$. We then compute critical exponents of the Ising model on higher-dimensional Sierpi\'{n}ski pyramids with various Hausdorff dimension via Monte-Carlo simulations and real-space RG analysis for $D\in[1,3]$. We find that only the correlation length exponent $\nu$ interpolates the values of integer-dimensional models. This implies that, contrary to a generally held belief, the universality class of quantum phase transition may not be uniquely determined by symmetry and spatial dimension of the system. This work initiates studies on quantum critical phenomena on graphs and networks which may be of significant importance in the context of quantum networks and communication.
\end{abstract}
\maketitle

\emph{Introduction.--} 
Quantum criticality is one of the most fascinating phenomena in many-body physics whose universal features often allow us to establish unexpected connection between different branches of physics. Yet, it is one of the most difficult phenomena to study because its analysis involves exponentially large Hilbert space due to diverging correlation length. In one dimension, powerful analytical tools, such as CFT and bosonization, along with efficient numerical techniques, such as DMRG and MPS, are available. Higher-dimensional quantum criticality, however, remains challenging to study with no established approaches though there are some attempts via entanglement renormalization~\cite{Vidal07, Evenbly09}.

\begin{figure}[htb!]
\centering
\includegraphics[width=1.00\linewidth]{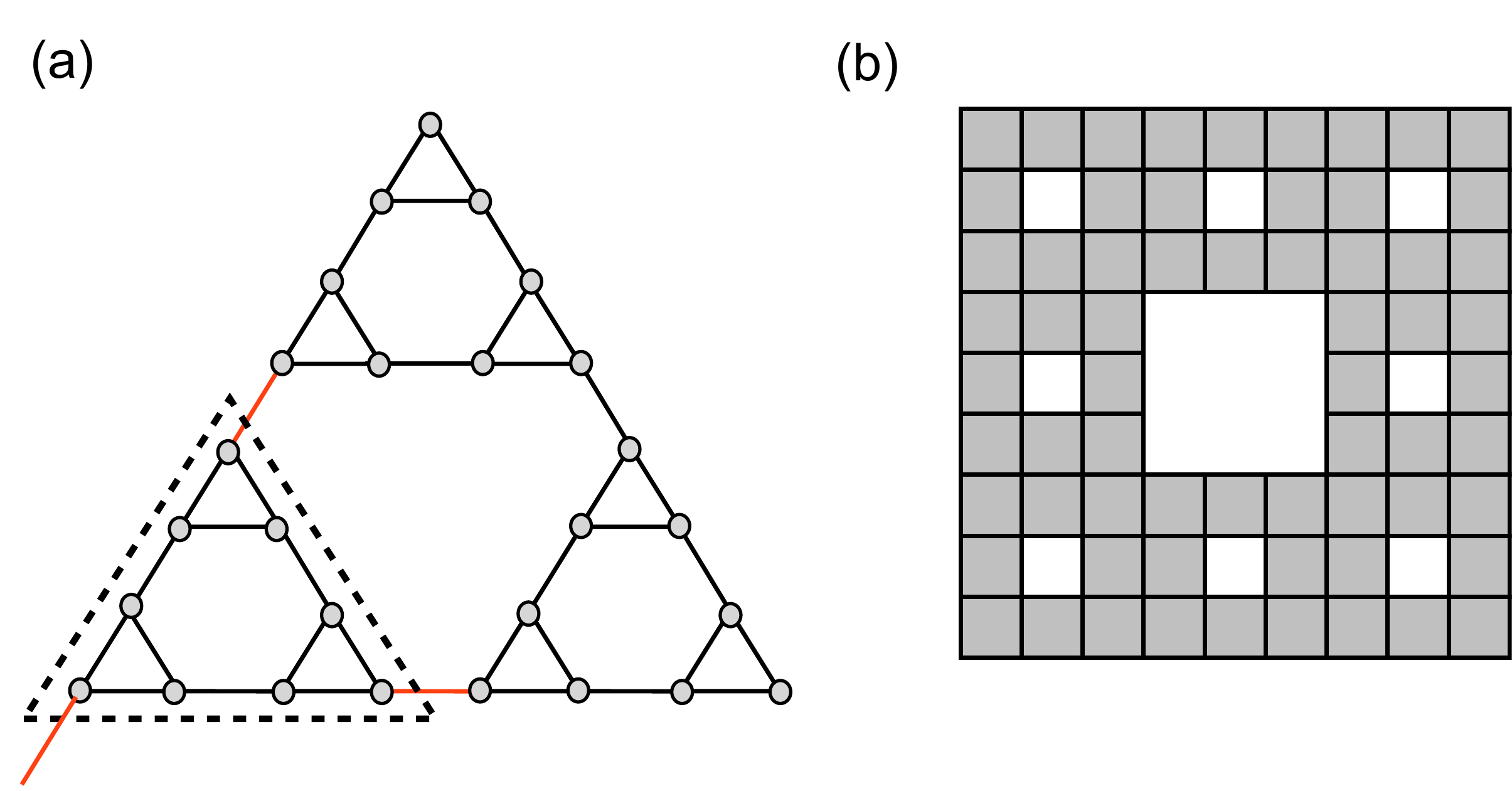}
\caption{ (a) The Sierpi\'{n}ski triangle with a finite Ramification number. Dotted lines above cut only three bonds. (b) The Sierpi\'{n}ski carpet with unbounded ramification number. 
} 
\label{Fig_lattice}
\end{figure}

In this paper, we revisit an approach of using \emph{fractal lattices} as a probe of higher-dimensional physics~\cite{Gefen80, Gefen81, Kaufman81, Domany83, Gefen83, Anacker87}. To be specific, consider a fractal lattice (Fig.~\ref{Fig_lattice}(a)), based on the Sierpi\'{n}ski triangle with Hausdorff dimension $D=\log 3/ \log 2 \approx 1.585$, on which some strongly interacting many-body system of our interest is prepared. The lattice has a finite ramification number, i.e. one can split it into disconnected parts by removing only finite number of bonds, and thus, one may efficiently simulate or analytically study it via renormalization group transformation. The hope is that quantum critical model on a fractal lattice has physical properties of a system with dimensionality $D>1$, but may be analyzed with computational complexity of $D=1$, serving as a physical realization of $\epsilon$-expansion. 

Despite its promising feature for unveiling $D>1$ physics, few models on ``efficiently simulable'' fractal lattices are currently known to exhibit critical phenomena~\cite{Tomczak96}, and much of previous work focuses on classical models. Thermal phase transitions of the classical Ising model on various fractal lattices have been exhaustively studied in the literature; yet, phase transitions were observed only on lattices with unbounded ramification number, such as the Sierpi\'{n}ski \emph{carpet} (Fig.~\ref{Fig_lattice}(b))~\cite{Gefen80}. Furthermore, critical exponents in the models on the carpet do not satisfy scaling relations, which would hold precisely if they were at quantum criticality~\cite{Gefen83, Monceau98}. Indeed, due to the violation of universality, a doubt on this approach has been raised for fractal lattices~\cite{Hu85}. Recently, the quantum Ising model on the Sierpi\'{n}ski carpet (Fig.~\ref{Fig_lattice}(b)) has been studied in~\cite{Yi13} where it was again found that critical exponents do not obey scaling relations. Indeed, whether quantum criticality may arise on fractal lattices is far from obvious as one often finds first-order transition on lattices with missing sites as in the dilute Ising model~\cite{Cardy_Text}. 

We demonstrate for the first time that fractal lattices can detect \emph{presence} or \emph{absence} of higher-dimensional quantum criticality in a many-body system by studying the quantum Ising and Potts models on the Sierpi\'{n}ski triangle and pyramid. As for \emph{presence}, we study the quantum Ising model on the Sierpi\'{n}ski triangle and observe quantum criticality with second-order transition where critical exponents satisfy scaling relations precisely. To our knowledge, this is the first time where quantum criticality and quantum phase transitions are observed on a fractal lattice. As for \emph{absence}, we study the quantum Potts model on the Sierpi\'{n}ski triangle and observe first-order phase transition. This is consistent with an estimate of critical dimension $D_{c}> 1.3$ above which phase transition is expected to be first-order~\cite{Nienhuis79, Nienhuis80}. We also study the quantum Ising model on three-dimensional Sierpi\'{n}ski \emph{pyramid} whose Hausdorff dimension happens to be \emph{exactly two}, and compare its critical exponents with those of two-dimensional quantum Ising model on a regular lattice.  

\begin{figure*}[htb!]
\centering
\includegraphics[width=0.95\linewidth]{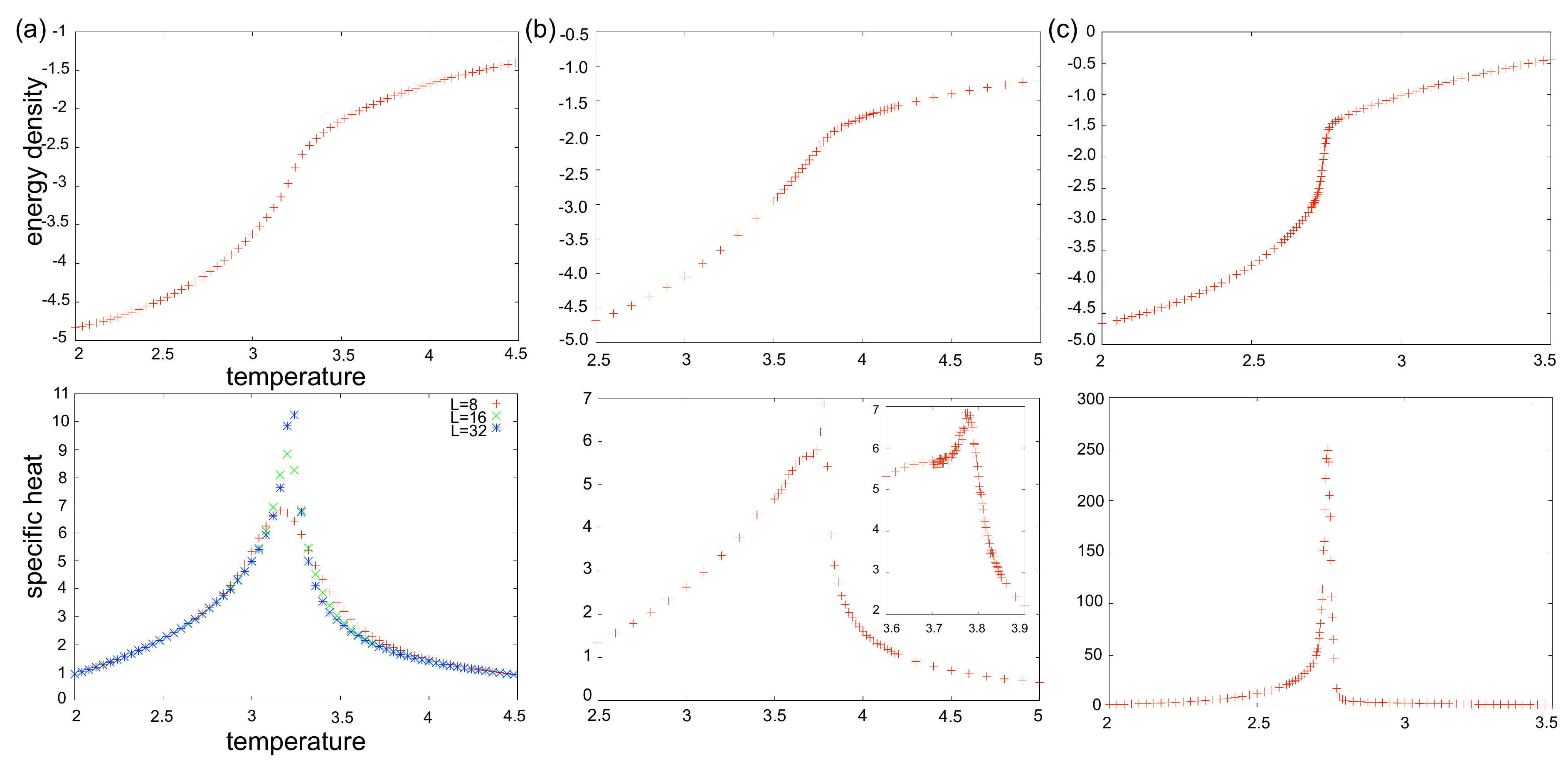}
\caption{(Color online) (a)(b) The quantum Ising model on the Sierpi\'{n}ski triangle and carpet. The inset shows behavior around $T_{c}$. (c) The quantum $3$-state Potts model on the Sierpi\'{n}ski triangle.
} 
\label{fig_data}
\end{figure*}

\emph{Sierpi\'{n}ski triangle.--}
We study the quantum Ising model in a transverse field on the Sierpi\'{n}ski triangle: 
\ba{
H(\epsilon) =  - (1-\epsilon) \sum_{(i,j)\in E} Z_{i}Z_{j} - \epsilon \sum_{j\in V} X_{j}
}
with $0\leq \epsilon \leq 1$ where $V$ ($E$) represents a set of vertices (edges) of the fractal. For $\epsilon=0$, the Hamiltonian is ferromagnetic with two ground states $|\psi_{0}\rangle = |0\rangle^{\otimes |V|}$ and $|\psi_{1}\rangle = |1\rangle^{\otimes |V|}$. For $\epsilon=1$, the Hamiltonian is paramagnetic with a unique ground state $|\phi\rangle = \big(\frac{1}{\sqrt{2}}(|0\rangle + |1\rangle)\big)^{\otimes |V|}$. Since the Hamiltonian connects two quantum phases with different symmetries, there must be a quantum phase transition (1st or 2nd) (see~\cite{Bravyi06} for instance).

The presence of quantum phase transition does not always imply the emergence of quantum criticality (second-order transition) with diverging correlation length. We perform Monte-Carlo simulations via standard Trotter-Suzuki treatment. The classical counterpart is defined on a lattice which is a stack of $L_{t}=L$ copies of the Sierpi\'{n}ski triangle where $L$ is the linear length of the original lattice. The effective spatial dimension is $d = \log 3 / \log 2 + 1 \approx 2.585$, and the Hamiltonian consists of Ising interactions between nearest-neighboring sites:
\ba{
H =  - J \sum_{k=1}^{L_{t}}\sum_{(i,j)\in E} Z_{i,k}Z_{j,k} - J \sum_{k=1}^{L_{t}} \sum_{j\in V} Z_{j,k}Z_{j,k+1}
}
where $Z_{i,k}$ acts on the $i$the spin in the $k$th layer. Simulations were performed by hybridizing local updates and cluster updates under periodic boundary conditions with thermalization and sampling steps of $10^{6}$. The energy density has no discrete jump, which is an indication of a second-order transition (Fig.~\ref{fig_data}(a)). 

\emph{Scaling relations.--}
Emergence of quantum criticality can be verified via scaling relations which can be derived from the following assumptions: (i) The system corresponds to a fixed point of RG transformation (scale-invariance). (ii) The correlation length $\xi$ diverges to $\sim L$. By checking scaling relations, one can verify above assumptions and confirm the emergence of quantum criticality. By standard finite-size scaling analysis, involving the Binder cumulant, we obtained 
\ba{
\alpha=0.034(3) \quad \nu = 0.76(1) \quad \frac{\beta}{\nu} = 0.237(3) \quad \frac{\gamma}{\nu} = 2.111(2).
}
Scaling relations relevant to us are 
\ba{
\nu d = 2- \alpha = 2 \beta + \gamma
}
which hold with good precision. One can even ``compute'' the effective spatial dimension from scaling relations, yielding $d\approx 2.585(8)$. Also, one has $\nu d + \alpha \approx 2.00(3)$. Thus, two assumptions above are correct, validating that the model exhibits $D>1$ quantum critical phenomena. 

The anomalous dimension $\eta$ can be computed from scaling relations. We obtain $\eta \approx - 0.111(2)$ which is negative. This is a characteristic of systems with singularities in the momentum space. The choice of $L_{t}=L$ in the Trotter-Suzuki mapping can be indirectly verified from the fact that scaling relations are precisely satisfied. This implies that the energy gap is $O(1/L)$ as in regular integer-dimensional quantum Ising models. This can be also verified by finding a universal function for the Binder cumulant. We also expect that the values of critical exponents are spatially isotropic despite the fact that the fractal lattice has spatially anisotropic structure. See Appendix for details.

For comparison, we have studied the quantum Ising model on the Sierpi\'{n}ski carpet (Fig.~\ref{Fig_lattice}(b)) which is $\log 8/\log 3$-dimensional~\cite{Yi13}. Under periodic boundary conditions, the energy density shows weak non-continuity, and peak values of the specific heat are different when approaching $T_{c}$ from the left and from the right (Fig.~\ref{fig_data}(b)). This implies (weak) first-order phase transition. It is surprising that quantum criticality arises only on the Sierpi\'{n}ski triangle, and not on the carpet despite the fact that only the former lattice is efficiently simulable.

\emph{Potts model.--}
Next, we demonstrate that fractal lattice models can predict the \emph{absence} of quantum criticality by studying the quantum Potts model on the Sierpi\'{n}ski triangle. We begin with a brief review of the classical $q$-state Potts model on a regular $d$-dimensional lattice: 
\ba{
H_{classical} = - J \sum_{(i,j)\in E} \delta(S_{i},S_{j})
}
where spin values are $S_{i}=0,\cdots,q-1$, and $\delta(S_{i},S_{j})=1$ for $S_{i}=S_{j}$ and $\delta(S_{i},S_{j})=0$ for $S_{i}\not=S_{j}$. For $q=2$, the Potts model is reduced to the Ising model. The Potts model serves as an interesting testbed for the validity of fractal lattice ideas for the following reason. For $q=3$, thermal transition is known to be second-order in $d=2$, but the transition becomes first-order in $d=3$. Thus there must be transition from second-order to first-order phase transition as dimension $d$ is varied. By treating $d$ as a continuous parameter, RG analysis predicts that the transition from ``2nd to 1st'' occurs around $d_{c}\approx 2.3$. 

We verify that the quantum Potts model on a fractal lattice can detect this transition by analyzing $\log 3/\log 2$-dimensional model. Via the Trotter-Suzuki mapping, its classical counterpart is $2.585$-dimensional, and should undergo a first-order transition. The energy density has a finite jump (Fig.~\ref{fig_data}(c)) which clearly indicates an emergence of a first-order transition which does not correspond to quantum criticality. This observation validates the use of fractal lattices for studying absence of quantum criticality, too.

\emph{Sierpi\'{n}ski pyramid.--}
It is generally believed that the universality class of quantum phase transition is determined solely by symmetry and spatial dimension. If so, the quantum Ising models on fractal lattices would interpolate integer-dimensional models. However, we find that this commonly held belief may not be true by studying the quantum Ising model on the Sierpi\'{n}ski pyramid, a generalization of the triangle to the three-dimensional space. Interestingly, its Hausdorff dimension is known to be \emph{exactly two}. For the pyramid model, we obtain
\ba{
\alpha= 0.02(1) \quad \nu = 0.660(5) \quad \frac{\beta}{\nu} = 0.304(6) \quad \frac{\gamma}{\nu} = 2.394(1)
}
while for the regular two-dimensional quantum Ising model
\ba{
\nu = 0.6301(4) \qquad \gamma = 1.2372(5).
}
The critical exponent $\nu$, related to divergence of correlation length via $\xi \sim |\lambda-\lambda_{c}|^{-\nu}$ where $\lambda$ is a parameter, is very close to the value in the regular quantum Ising model while other critical exponents are different. This implies that the universality class may not be uniquely determined by symmetry and spatial dimension of the system. This result may provide useful insight into the conformal bootstrapping~\cite{El-Showk12}. 

\begin{figure}[htb!]
\centering
\includegraphics[width=1.00\linewidth]{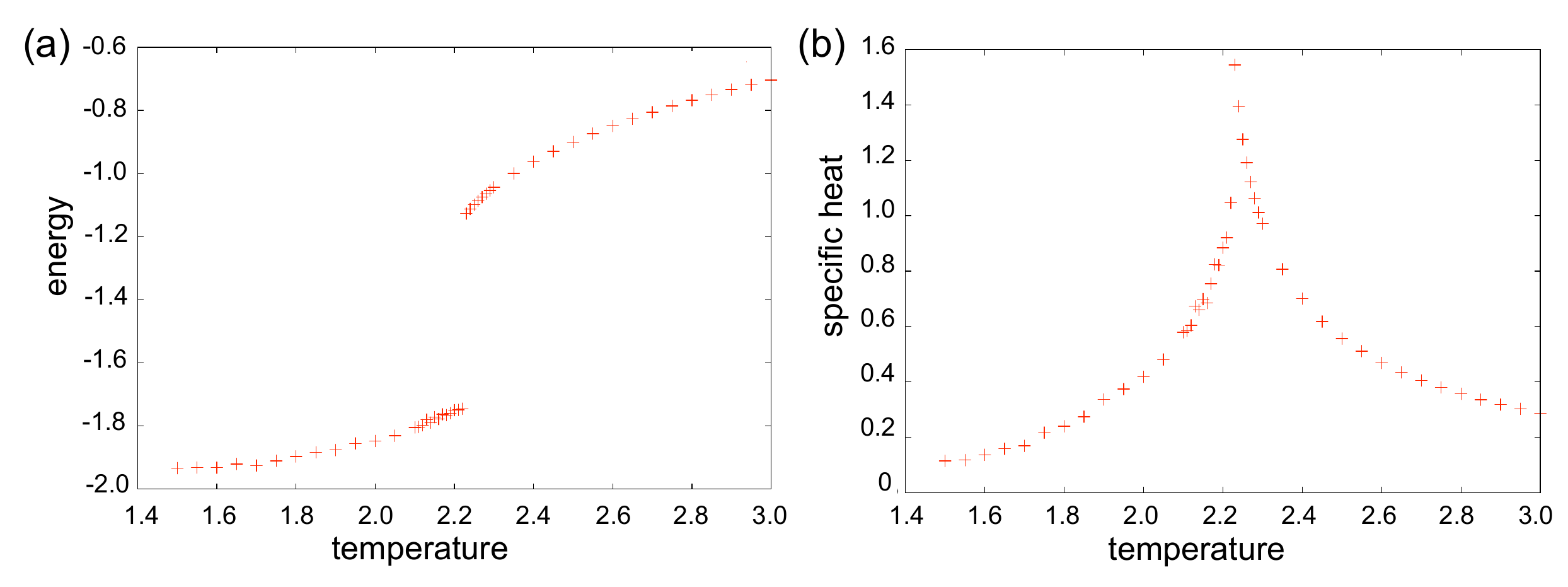}
\caption{(Color online) Classical fractal code. (a) Energy density. (c) Specific heat. 
} 
\label{fig_data2}
\end{figure}

\emph{Fractal code.--}
Another interesting spin models where fractal structures emerge are classical and quantum fractal codes~\cite{Newman99, Beni11b,Haah11,Beni13} which are governed by translation symmetric Hamiltonians on regular lattices, but have a large number of degenerate ground states whose spin configurations are fractal-like. Consider the Hamiltonian on a regular two-dimensional square lattice with $L\times L$ sites ($L=2^{m}$)
\ba{
H=-\sum_{i=1}^{L}\sum_{j=2}^{L}Z_{i,j}Z_{i+1,j}Z_{i+1,j+1}
}
with periodic boundary conditions in $\hat{x}$-direction and open boundary conditions in $\hat{y}$-direction. This Hamiltonian has the following fractal-like ground state (for $L=2^3$):
\ba{
\psi = \begin{bmatrix}
1 & 0 & 0 & 0 & 0 & 0 & 0 & 0 \\
1 & 1 & 0 & 0 & 0 & 0 & 0 & 0 \\
1 & 0 & 1 & 0 & 0 & 0 & 0 & 0 \\
1 & 1 & 1 & 1 & 0 & 0 & 0 & 0 \\
1 & 0 & 0 & 0 & 1 & 0 & 0 & 0 \\
1 & 1 & 0 & 0 & 1 & 1 & 0 & 0 \\
1 & 0 & 1 & 0 & 1 & 0 & 1 & 0 \\
1 & 1 & 1 & 1 & 1 & 1 & 1 & 1 
\end{bmatrix}
}
where spin values are represented in the computational basis $0, 1$. Upper-left corner corresponds to a spin at the position $(1,1)$ and lower-right corner corresponds to a spin at $(L,L)$. A family of these classical fractal codes are known to saturate theoretical limits on classical information storage capacity of local Hamiltonians~\cite{Beni11b}, and corresponds to limit cycles of real-space wavefunction renormalization instead of fixed points~\cite{Beni13}. 

Here, we consider quantum phase transition driven by transverse field:
\ba{
H(\epsilon)=(1-\epsilon)\sum_{i,j}Z_{i,j}Z_{i+1,j}Z_{i+1,j+1} - \epsilon \sum_{i,j}X_{i,j}
}
via the Trotter-Suzuki mapping. Following an argument in~\cite{Beni10b}, there must be a quantum phase transition due to the Lieb-Robinson bound~\cite{Bravyi06}. From self-duality of the model, the transition is at $\epsilon=1/2$. We observe evident first-order transition signatures as depicted in Fig.~\ref{fig_data2}. This is due to the fact that the Hamiltonian $H(0)$ has a diverging number of degenerate ground states, as well as a lot of local energy minima. 

\emph{Real-space RG analysis.--}
Finally, we perform real-space RG analysis of the quantum Ising model on the generalized Sierpi\'{n}ski pyramid. In $m$-dimensional space, the Hausdorff dimension is $\log (m+1)/ \log 2$. Unlike regular integer-dimensional lattices, the ramification number of the generalized Sierpi\'{n}ski triangle is finite. Thus there is no unwanted extra term arising in renormalized Hamiltonian which would be present for lattices with diverging ramification number. 

\begin{figure}[htb!]
\centering
\includegraphics[width=1.0\linewidth]{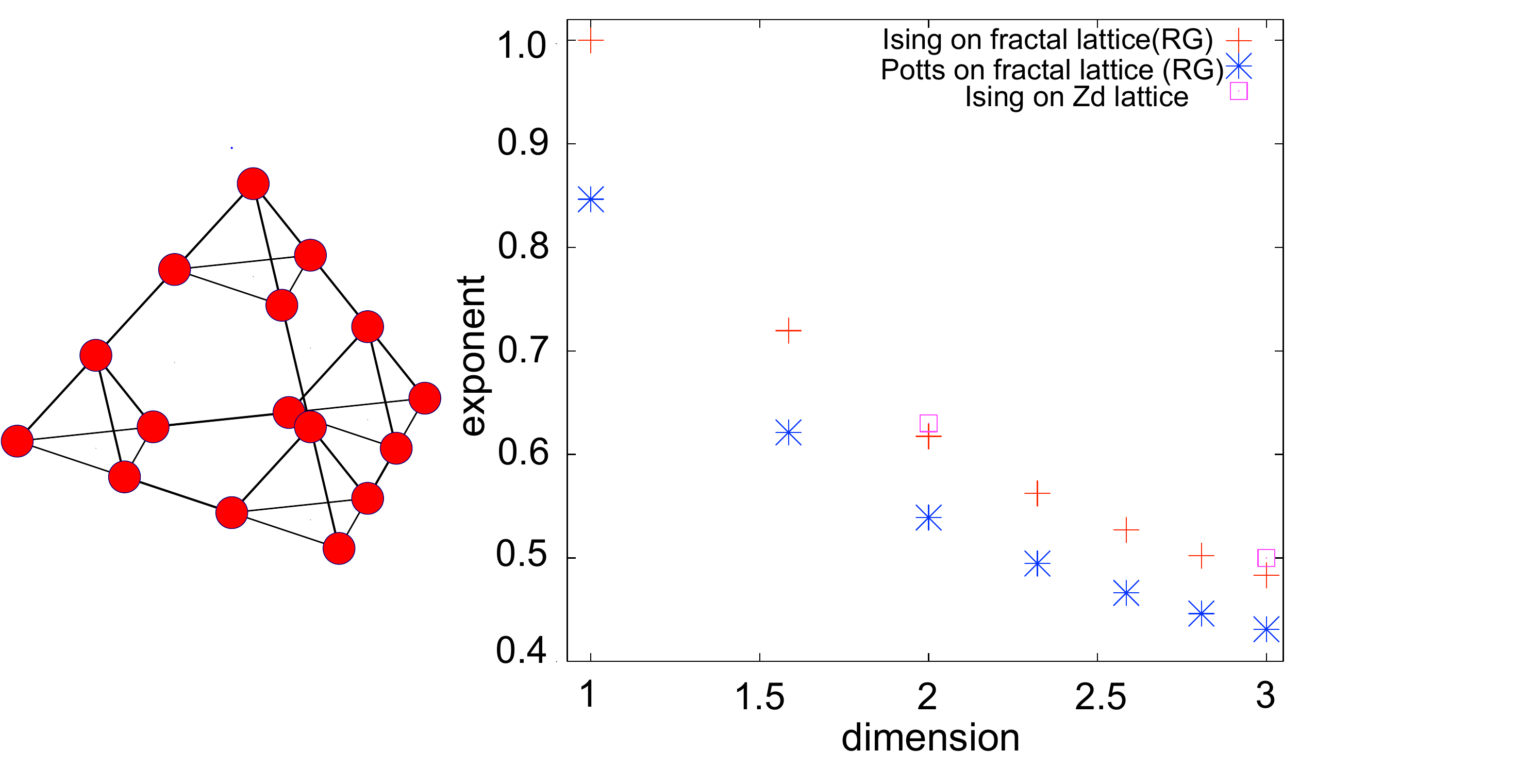}
\caption{(Color online) (a) The Sierpi\'{n}ski pyramid. (b) The correlation length exponent of the quantum Ising and Potts models from real-space RG. 
} 
\label{fig_high}
\end{figure}

We employ real-space RG scheme, first introduced by Fernandez-Pacheco~\cite{Fernandez-Pacheco79}, which is able to predict $\nu$ with surprisingly good precision (around $2- 3$ percent). The details are presented in an accompanying paper~\cite{Kubica14}. We obtain $\nu=0.72,0.62$ for $D=\log 3/ \log 2, \log 4/ \log 2$ which is in good agreement with numerical estimates. As can be seen in Fig.~\ref{fig_high}, fractal lattice models interpolate integer-dimensional models, which implies that $\nu$ is mostly determined by symmetry and spatial dimension of the system. 

\emph{Discussion.--} 
We have demonstrated that fractal lattices can detect the emergence or absence of higher-dimensional quantum criticality. We have also found that, contrary to a generally held belief, symmetry and spatial dimension of the system do not determine the universality class completely for $D\geq2$. While our study is limited to the quantum Ising (Potts) model, this work may give confidence and validation of the use of fractal lattices for studying higher-dimensional critical phenomena and lead to systematic study of various many-body physics on fractal lattices. We have opted for using a traditional Monte-Carlo method in order to highlight the difference between classical and quantum Ising models on fractal lattices. It will be interesting to apply modern numerical methods such as entanglement renormalization~\cite{Vidal07} to our models. 

Classical statistical physics on various graphs and networks is a subject of significant interest, bridging various fields of physics, biology and social sciences~\cite{Albert02}. Progresses of experimental techniques will eventually lead to realizations of quantum many-body networks in an isolated system at zero temperature. Our work serves as a stepping stone toward studies of quantum statistical physics on various graphs and networks, which will be of significant interest and importance in the context of quantum networks and communication. 

\emph{Ackowledgment --}
We thank Glen Evenbly and Shintaro Takayoshi for helpful discussion and comments. BY is supported by the David and Ellen Lee Postdoctoral fellowship. We acknowledge funding provided by the Institute for Quantum Information and Matter, an NSF Physics Frontiers Center with support of the Gordon and Betty Moore Foundation (Grants No. PHY-0803371 and PHY-1125565). This work used the Extreme Science and Engineering Discovery Environment (XSEDE), which is supported by National Science Foundation grant number OCI-1053575.

\appendix

\section{Further discussion on numerical data}

In the appendix, we present details of numerical methods and supplementary numerical results. 

\begin{figure}[htb!]
\centering
\includegraphics[width=0.95\linewidth]{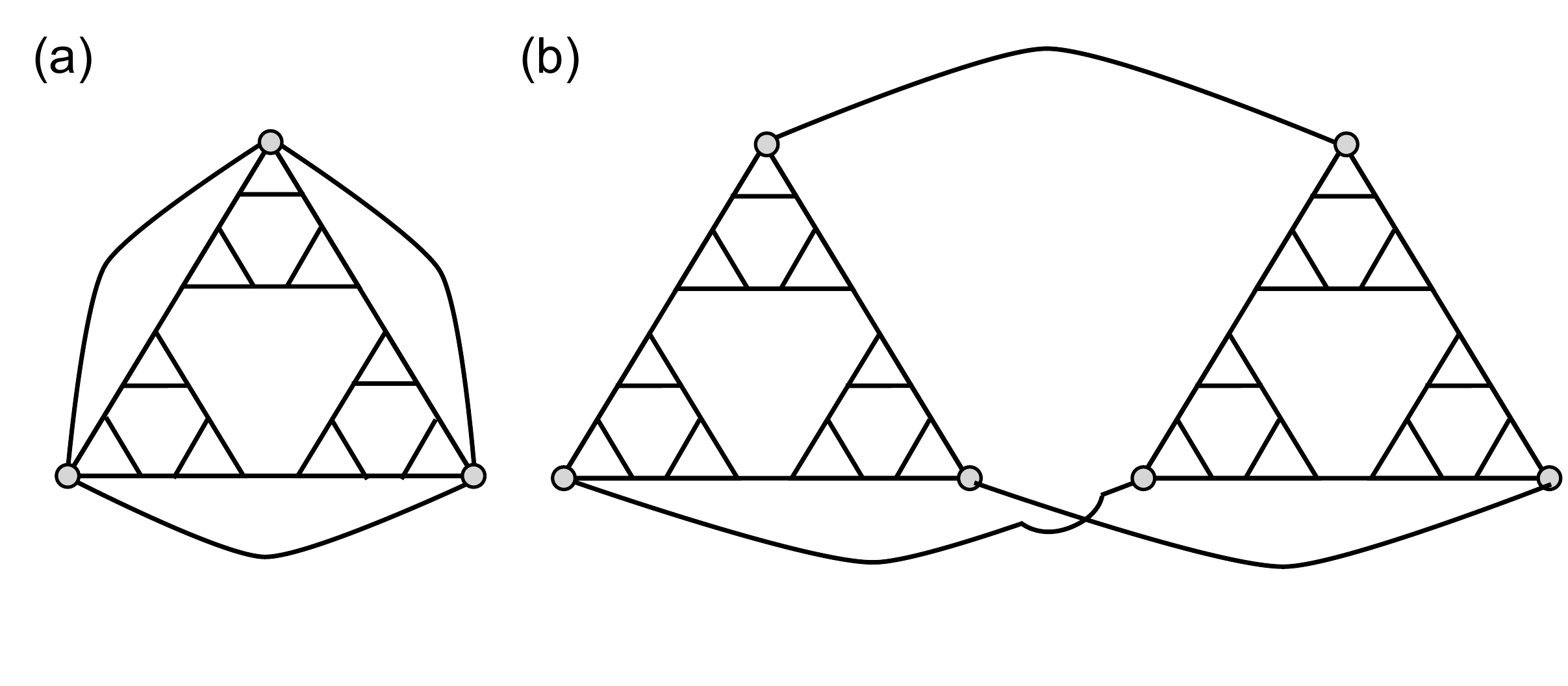}
\caption{Two types of boundary conditions.
} 
\label{fig_PBC}
\end{figure}

\emph{Boundary conditions.--} In simulating the quantum Ising model on the Sierpi\'{n}ski triangle, we tried two types of periodic boundary conditions (see Fig.~\ref{fig_PBC}). (a) Three corner sites of the lattice are coupled with Ising couplings. (b) Two copies of the Sierpi\'{n}ski triangle are prepared, and corners from each triangle are coupled. We found that the boundary condition (b) has significantly smaller statistical errors with good convergence and is suitable for our purposes, and thus, our results are based on the boundary condition (b). 

\emph{Critical exponents.--}
We briefly summarize a standard method of estimating critical exponents via finite-size scaling method. For detailed descriptions, readers are referred to standard textbooks~\cite{Cardy_Text}. One can estimate the value of $\gamma/\nu$ by analyzing the divergence of peak values of magnetic susceptibilities for different $L$. In particular, one has
\ba{
\chi_{peak} \sim L^{\gamma/\nu}
}
where $\chi_{peak}$ is the peak value of magnetic susceptibility. One can estimate the transition temperature $T_{c}$ by finding the crossing point of the Binder cumulants $U^{(4)}$ for different $L$:
\ba{
U^{(4)}=1 - \frac{\langle m^4 \rangle}{3\langle m^2 \rangle^2}.
}
One can estimate the value of $\beta/\nu$ by analyzing the decay of magnetization at the critical temperature $T_{c}$ for different $L$:
\ba{
m(T=T_{c}) \sim L^{- \beta/\nu}.
}
One can estimate $\nu$ by finding a universal function for the Binder cumulant:
\ba{
U^{(4)} = f(tL^{1/\nu})
}
where $t=(T-T_{c})/T_{c}$ is the normalized temperature.

\begin{figure*}[htb!]
\centering
\includegraphics[width=0.80\linewidth]{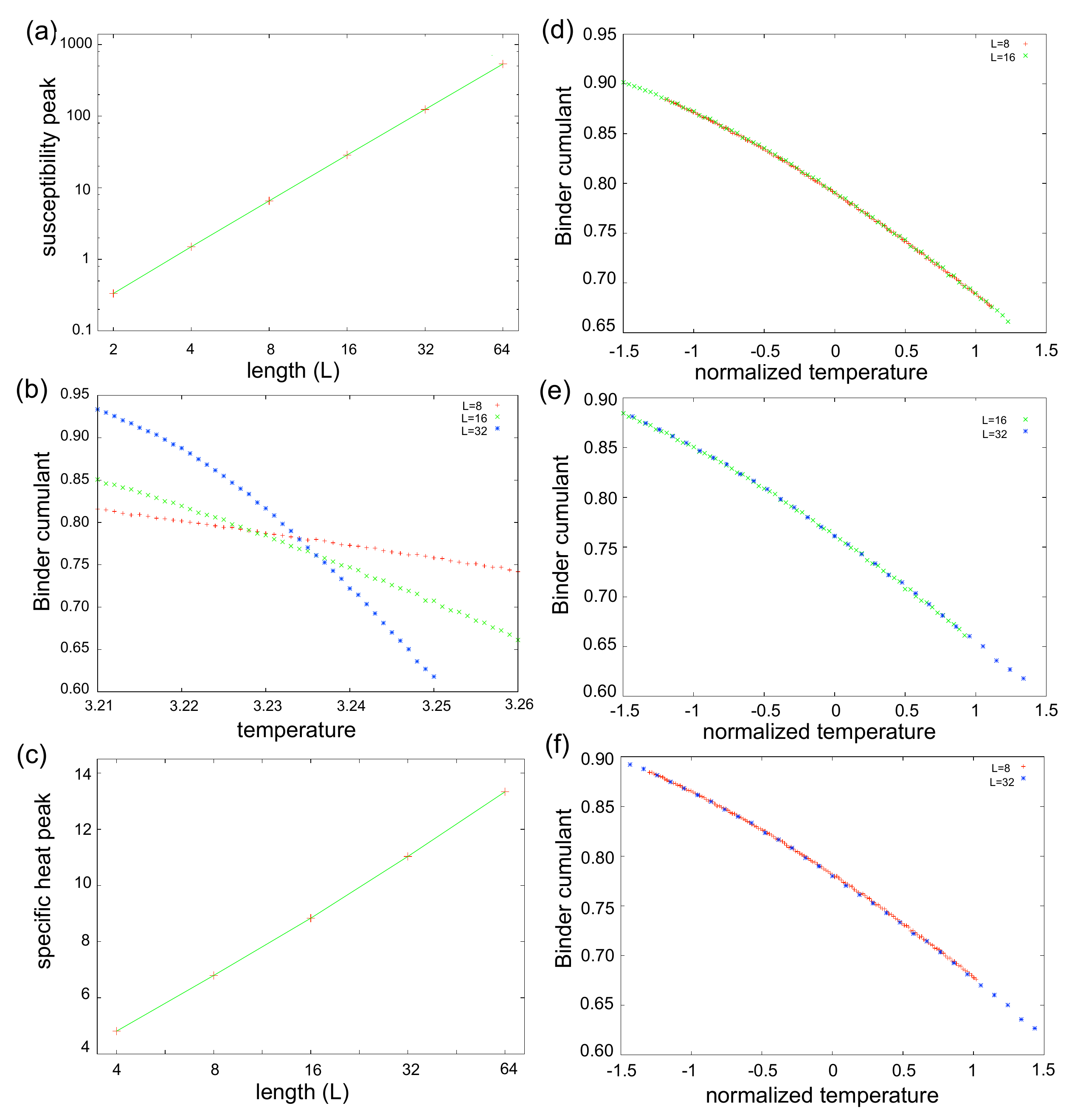}
\caption{(Color online) Numerical results for the Sierpi\'{n}ski triangle. (a) Peaks of susceptibilities. (b) Crossing points of the Binder cumulants. (c) Peaks of specific heat. (d)(e)(f) Binder universal function ($U^{(4)}$ vs $tL^{1/\nu}$ for $\nu=0.76$). 
} 
\label{fig_data1}
\end{figure*}

Below, we describe our analysis for the Sierpi\'{n}ski triangle below. As for the estimate of $\gamma/\nu$, peak values of magnetic susceptibility are plotted in Fig.~\ref{fig_data1}(a). We find that the slope of the fit in a log-log graph is
\ba{
\frac{\gamma}{\nu}=2.111(1).
}
As for the transition temperature $T_{c}$, crossing points of the Binder cumulants for $L=8,16,32$ are shown in Fig.~\ref{fig_data1}(b). We find $T_{c}(8,16) =3.228$, $T_{c}(16,32) =3.236$ and $T_{c}(8,32) =3.234$ where $T_{c}(L_{1},L_{2})$ represents a crossing point for $L=L_{1}$ and $L=L_{2}$. Since estimations of $\beta/\nu$ and $\nu$ heavily depend on the locations of $T_{c}$, we estimate those values at each crossing point. We find $\frac{\beta}{\nu}(8,16) = 0.234$, $\frac{\beta}{\nu}(16,32) = 0.239$ and $\frac{\beta}{\nu}(8,32) =  0.239$. Thus, we obtain 
\ba{
\frac{\beta}{\nu} = 0.237(3)
}
by taking their average. We also find $\nu(8,16) = \nu(16,32) = \nu(8,32) =  0.76(1)$ where the Binder cumulants $U^{(4)}$ as a function of $tL^{1/\nu}$ for three crossing points are shown in Fig.~\ref{fig_data1}(d)(e)(f). The results are summarized below.

\begin{table}[h]
\centering
\begin{tabular}{c|c|c|c}
  & \ \  (8,16) \ \ & \ \ (16,32) \  \ & \ \ (8,32) \ \ \\
\hline
$T_{c}$ & $3.228$ & $3.236$ & $3.234$ \\
$\beta/\nu$ & $0.234$ & $0.239$ & $0.239$  \\
$\nu$ & $0.76(1)$ & $0.76(1)$ & $0.76(1)$ \\
\end{tabular}
\end{table}

For Ising-type spin models, estimation of $\alpha$ via Monte-Carlo simulation is generally hard since $\alpha$ is small. The peak values of $\alpha$ show logarithmic growth for both the Sierpi\'{n}ski triangle and pyramid, which indicates $\alpha \sim 0$ as depicted in Fig.~\ref{fig_data1}(c). Yet, we observe a slight signature of a polynomial divergence in peak values of $\alpha$. By fitting $C_{peak}$ with
\ba{
(a\log L + b)L^{\alpha}
}
for some constant $a,b$, we obtain
\ba{
\alpha \approx 0.034(3). 
}
With this estimate, we recover the scaling relation $\nu d = 2-\alpha$ with a good precision.   

\emph{Anisotropy.--} 
We have studied the classical Ising model which arises from the quantum Ising model on a fractal lattice via Trotter-Suzuki method. This classical counterpart has spatially anisotropic structure and may possess two different values of the correlation length and critical exponents in the imaginary time direction and inside the layer of the fractal lattice. Our numerical estimation of critical exponents is based on the finite-size scaling method, which implicitly assumes that the values of critical exponents are spatially isotropic. 

In order to test the validity of this assumption, we have evaluated two-point correlation function in the imaginary time direction and on the single layer of the fractal lattice. We have defined the distance between two spins on a fractal lattice as Euclidean distance in the space where the fractal lattice is embedded (i.e. for the Sierpi\'{n}ski triangle, it is $\mathbb{R}^{2}$). We find that, for small $L$, the transition temperatures $T_{time}$ and $T_{layer}$ for two cases are slightly different. As $L$ increases, the separation between $T_{time}$ and $T_{layer}$ seems to decrease. We obtain rough estimations of the values of the anomalous scaling dimension $\eta$ by fitting two-point correlation function with $e^{-r/\xi}\cdot r^{-d+2-\eta}$ where $d$ is the Hausdorff dimension of the lattice on which the classical model is defined. Our estimations of $\eta$ from two-point correlation function are consistent with the one from the finite-size scaling method. 

Below, we describe our findings for $L=16$. The estimation of the critical temperature $T_{c}$ from the finite-size scaling method is $T_{c}\approx 3.23$. For $T \in [3.22,3.29]$, the value of the correlation length $\xi$ is $10^2-10^6$. In such a temperature regime, we fit the correlation function simply with $r^{-d+2-\eta}$. We then find a narrower temperature regime where the fitting error becomes significantly small from which we estimate the values of $T_{time}$ and $T_{layer}$. For $L=16$, we find $T_{time}\approx 3.27(1)$ and $T_{layer}\approx 3.23(1)$. We also find that $\eta_{time}\approx - 0.20$ and $\eta_{layer}\approx -0.10$ at $T=T_{time}$, and $\eta_{time}\approx -0.08$ and $\eta_{layer}\approx +0.04$ at $T=T_{layer}$. We observe that $\eta_{time}$ at $T=T_{time}$ and $\eta_{layer}$ at $T=T_{layer}$ are close and consistent with an estimate from the finite-size scaling method, $\eta \approx 0.11$. We then repeat a similar analysis for $L=32$. We find that $\eta_{time}\approx - 0.09$ at $T_{time}\approx 3.233(2)$ and $\eta_{layer}\approx - 0.12$ at $T_{layer}\approx 3.245(2)$. We observe that the separation between $T_{time}$ and $T_{layer}$ decreases and estimations of $\eta$ from two-point correlation function are consistent with the one from the finite-size scaling method. 

With these observations, we expect that the spatial anisotropy will disappear at $L \rightarrow \infty$. However, it should be emphasized that it is rather challenging to obtain precise estimations of $T_{time}$, $T_{layer}$ and $\eta_{time}$, $\eta_{layer}$ from two-point correlation function. Thus, these observations are not enough to rule out the possibility of spatially anisotropic critical exponents. 

\emph{Energy gap.--} 
Our numerical estimation of critical exponents is based on Monte-Carlo simulations for $L_{t}=L$ in the Trotter-Suzuki method. The choice of $L_{t}=L$ is valid when the energy gap at the criticality is $O(1/L)$. This is normally true for systems that undergo second-order quantum phase transition. Also the fact that scaling relations are precisely satisfied gives an indirect verification of the choice of $L_{t}=L$. Yet, we do not know the scaling of the energy gap \emph{a priori}. 

Let us assume that the energy gap scales as $L^{-z}$ at the transition point. The most convincing way of finding $z$ would be to perform exact diagonalization, which is computationally intractable for two-dimensional quantum systems. Exact diagonalization of two-dimensional quantum Ising model has been performed for a system with $6 \times 6$ spins in~\cite{Hamer00} where the prediction of $z=1$ has been verified. However, the drawback is that exact diagonalization is possible only for systems with small number of spins. Instead, one can estimate the value of $z$ by looking for the universal form of the Binder cumulant (see~\cite{Guo94} for instance)
\ba{
U^{(4)} = g(t L^{1/\nu},L_{t}L^{-z}).
}
In order to find the value of $z$, one needs to perform Monte-Carlo simulations with $L_{t}=L^{a}$ for various values of $a$. If $a=z$, one is able to find the value of $\nu$ such that the Binder cumulant has a universal form. We have seen that, with $a=1$, one can find a universal form, which implies $z=1$. For comparison, we have performed Monte-Carlo simulations for $a=\log 3/\log 2, \log 4/ \log 2$. In particular, for $L=2^{m}$, we have studied numerical results with $L_{t}=3^{m},4^{m}$ for $m=2,3,4$. In both cases, there is no choice of $\nu$ which leads to a universal form of the Binder cumulant. We also find that the discrepancies among different curves seem to increase as $a$ increases.


\begin{thebibliography}{27}
\expandafter\ifx\csname natexlab\endcsname\relax\def\natexlab#1{#1}\fi
\expandafter\ifx\csname bibnamefont\endcsname\relax
  \def\bibnamefont#1{#1}\fi
\expandafter\ifx\csname bibfnamefont\endcsname\relax
  \def\bibfnamefont#1{#1}\fi
\expandafter\ifx\csname citenamefont\endcsname\relax
  \def\citenamefont#1{#1}\fi
\expandafter\ifx\csname url\endcsname\relax
  \def\url#1{\texttt{#1}}\fi
\expandafter\ifx\csname urlprefix\endcsname\relax\def\urlprefix{URL }\fi
\providecommand{\bibinfo}[2]{#2}
\providecommand{\eprint}[2][]{\url{#2}}

\bibitem[{\citenamefont{Vidal}(2007)}]{Vidal07}
\bibinfo{author}{\bibfnamefont{G.}~\bibnamefont{Vidal}},
  \bibinfo{journal}{Phys. Rev. Lett.} \textbf{\bibinfo{volume}{99}},
  \bibinfo{pages}{220405} (\bibinfo{year}{2007}).

\bibitem[{\citenamefont{Evenbly and Vidal}(2009)}]{Evenbly09}
\bibinfo{author}{\bibfnamefont{G.}~\bibnamefont{Evenbly}} \bibnamefont{and}
  \bibinfo{author}{\bibfnamefont{G.}~\bibnamefont{Vidal}},
  \bibinfo{journal}{Phys. Rev. Lett.} \textbf{\bibinfo{volume}{102}},
  \bibinfo{pages}{180406} (\bibinfo{year}{2009}).

\bibitem[{\citenamefont{Gefen et~al.}(1980)\citenamefont{Gefen, Mandelbrot, and
  Aharony}}]{Gefen80}
\bibinfo{author}{\bibfnamefont{Y.}~\bibnamefont{Gefen}},
  \bibinfo{author}{\bibfnamefont{B.~B.} \bibnamefont{Mandelbrot}},
  \bibnamefont{and} \bibinfo{author}{\bibfnamefont{A.}~\bibnamefont{Aharony}},
  \bibinfo{journal}{Phys. Rev. Lett.} \textbf{\bibinfo{volume}{45}},
  \bibinfo{pages}{855} (\bibinfo{year}{1980}).

\bibitem[{\citenamefont{Gefen et~al.}(1981)\citenamefont{Gefen, Aharony,
  Mandelbrot, and Kirkpatrick}}]{Gefen81}
\bibinfo{author}{\bibfnamefont{Y.}~\bibnamefont{Gefen}},
  \bibinfo{author}{\bibfnamefont{A.}~\bibnamefont{Aharony}},
  \bibinfo{author}{\bibfnamefont{B.~B.} \bibnamefont{Mandelbrot}},
  \bibnamefont{and}
  \bibinfo{author}{\bibfnamefont{S.}~\bibnamefont{Kirkpatrick}},
  \bibinfo{journal}{Phys. Rev. Lett.} \textbf{\bibinfo{volume}{47}},
  \bibinfo{pages}{1771} (\bibinfo{year}{1981}).

\bibitem[{\citenamefont{Kaufman and Griffiths}(1981)}]{Kaufman81}
\bibinfo{author}{\bibfnamefont{M.}~\bibnamefont{Kaufman}} \bibnamefont{and}
  \bibinfo{author}{\bibfnamefont{R.~B.} \bibnamefont{Griffiths}},
  \bibinfo{journal}{Phys. Rev. B} \textbf{\bibinfo{volume}{24}},
  \bibinfo{pages}{496} (\bibinfo{year}{1981}).

\bibitem[{\citenamefont{Domany et~al.}(1983)\citenamefont{Domany, Alexander,
  Bensimon, and Kadanoff}}]{Domany83}
\bibinfo{author}{\bibfnamefont{E.}~\bibnamefont{Domany}},
  \bibinfo{author}{\bibfnamefont{S.}~\bibnamefont{Alexander}},
  \bibinfo{author}{\bibfnamefont{D.}~\bibnamefont{Bensimon}}, \bibnamefont{and}
  \bibinfo{author}{\bibfnamefont{L.~P.} \bibnamefont{Kadanoff}},
  \bibinfo{journal}{Phys. Rev. B} \textbf{\bibinfo{volume}{28}},
  \bibinfo{pages}{3110} (\bibinfo{year}{1983}).

\bibitem[{\citenamefont{Gefen et~al.}(1983)\citenamefont{Gefen, Meir,
  Mandelbrot, and Aharony}}]{Gefen83}
\bibinfo{author}{\bibfnamefont{Y.}~\bibnamefont{Gefen}},
  \bibinfo{author}{\bibfnamefont{Y.}~\bibnamefont{Meir}},
  \bibinfo{author}{\bibfnamefont{B.~B.} \bibnamefont{Mandelbrot}},
  \bibnamefont{and} \bibinfo{author}{\bibfnamefont{A.}~\bibnamefont{Aharony}},
  \bibinfo{journal}{Phys. Rev. Lett.} \textbf{\bibinfo{volume}{50}},
  \bibinfo{pages}{145} (\bibinfo{year}{1983}).

\bibitem[{\citenamefont{Anacker and Kopelman}(1987)}]{Anacker87}
\bibinfo{author}{\bibfnamefont{L.~W.} \bibnamefont{Anacker}} \bibnamefont{and}
  \bibinfo{author}{\bibfnamefont{R.}~\bibnamefont{Kopelman}},
  \bibinfo{journal}{Phys. Rev. Lett.} \textbf{\bibinfo{volume}{58}},
  \bibinfo{pages}{289} (\bibinfo{year}{1987}).

\bibitem[{\citenamefont{Tomczak}(1996)}]{Tomczak96}
\bibinfo{author}{\bibfnamefont{P.}~\bibnamefont{Tomczak}},
  \bibinfo{journal}{Phys. Rev. B} \textbf{\bibinfo{volume}{53}},
  \bibinfo{pages}{R500} (\bibinfo{year}{1996}).

\bibitem[{\citenamefont{Monceau et~al.}(1998)\citenamefont{Monceau, Perreau,
  and H{\'e}bert}}]{Monceau98}
\bibinfo{author}{\bibfnamefont{P.}~\bibnamefont{Monceau}},
  \bibinfo{author}{\bibfnamefont{M.}~\bibnamefont{Perreau}}, \bibnamefont{and}
  \bibinfo{author}{\bibfnamefont{F.}~\bibnamefont{H{\'e}bert}},
  \bibinfo{journal}{Phys. Rev. B} \textbf{\bibinfo{volume}{58}},
  \bibinfo{pages}{6386} (\bibinfo{year}{1998}).

\bibitem[{\citenamefont{Hu}(1985)}]{Hu85}
\bibinfo{author}{\bibfnamefont{B.}~\bibnamefont{Hu}}, \bibinfo{journal}{Phys.
  Rev. Lett.} \textbf{\bibinfo{volume}{55}}, \bibinfo{pages}{2316}
  (\bibinfo{year}{1985}).

\bibitem[{\citenamefont{Yi}(2013)}]{Yi13}
\bibinfo{author}{\bibfnamefont{H.}~\bibnamefont{Yi}}, \bibinfo{journal}{Phys.
  Rev. E} \textbf{\bibinfo{volume}{88}}, \bibinfo{pages}{014105}
  (\bibinfo{year}{2013}).

\bibitem[{\citenamefont{{Cardy}}(1996)}]{Cardy_Text}
\bibinfo{author}{\bibfnamefont{J.}~\bibnamefont{{Cardy}}},
  \emph{\bibinfo{title}{Scaling and Renormalization in Statistical Physics}}
  (\bibinfo{publisher}{Cambridge University Press, Cambridge},
  \bibinfo{year}{1996}).

\bibitem[{\citenamefont{Nienhuis et~al.}(1979)\citenamefont{Nienhuis, Berker,
  Riedel, and Schick}}]{Nienhuis79}
\bibinfo{author}{\bibfnamefont{B.}~\bibnamefont{Nienhuis}},
  \bibinfo{author}{\bibfnamefont{A.~N.} \bibnamefont{Berker}},
  \bibinfo{author}{\bibfnamefont{E.~K.} \bibnamefont{Riedel}},
  \bibnamefont{and} \bibinfo{author}{\bibfnamefont{M.}~\bibnamefont{Schick}},
  \bibinfo{journal}{Phys. Rev. Lett.} \textbf{\bibinfo{volume}{43}},
  \bibinfo{pages}{737} (\bibinfo{year}{1979}).

\bibitem[{\citenamefont{Nienhuis et~al.}(1980)\citenamefont{Nienhuis, Riedel,
  and Schick}}]{Nienhuis80}
\bibinfo{author}{\bibfnamefont{B.}~\bibnamefont{Nienhuis}},
  \bibinfo{author}{\bibfnamefont{E.~K.} \bibnamefont{Riedel}},
  \bibnamefont{and} \bibinfo{author}{\bibfnamefont{M.}~\bibnamefont{Schick}},
  \bibinfo{journal}{Journal of Physics A: Mathematical and General}
  \textbf{\bibinfo{volume}{13}}, \bibinfo{pages}{L189} (\bibinfo{year}{1980}).

\bibitem[{\citenamefont{Bravyi et~al.}(2006)\citenamefont{Bravyi, Hastings, and
  Verstraete}}]{Bravyi06}
\bibinfo{author}{\bibfnamefont{S.}~\bibnamefont{Bravyi}},
  \bibinfo{author}{\bibfnamefont{M.~B.} \bibnamefont{Hastings}},
  \bibnamefont{and}
  \bibinfo{author}{\bibfnamefont{F.}~\bibnamefont{Verstraete}},
  \bibinfo{journal}{Phys. Rev. Lett.} \textbf{\bibinfo{volume}{97}},
  \bibinfo{pages}{050401} (\bibinfo{year}{2006}).

\bibitem[{\citenamefont{El-Showk et~al.}(2012)\citenamefont{El-Showk, Paulos,
  Poland, Rychkov, Simmons-Duffin, and Vichi}}]{El-Showk12}
\bibinfo{author}{\bibfnamefont{S.}~\bibnamefont{El-Showk}},
  \bibinfo{author}{\bibfnamefont{M.~F.} \bibnamefont{Paulos}},
  \bibinfo{author}{\bibfnamefont{D.}~\bibnamefont{Poland}},
  \bibinfo{author}{\bibfnamefont{S.}~\bibnamefont{Rychkov}},
  \bibinfo{author}{\bibfnamefont{D.}~\bibnamefont{Simmons-Duffin}},
  \bibnamefont{and} \bibinfo{author}{\bibfnamefont{A.}~\bibnamefont{Vichi}},
  \bibinfo{journal}{Phys. Rev. D} \textbf{\bibinfo{volume}{86}},
  \bibinfo{pages}{025022} (\bibinfo{year}{2012}).

\bibitem[{\citenamefont{Newman and Moore}(1999)}]{Newman99}
\bibinfo{author}{\bibfnamefont{M.~E.~J.} \bibnamefont{Newman}}
  \bibnamefont{and} \bibinfo{author}{\bibfnamefont{C.}~\bibnamefont{Moore}},
  \bibinfo{journal}{Phys. Rev. E} \textbf{\bibinfo{volume}{60}},
  \bibinfo{pages}{5068} (\bibinfo{year}{1999}).

\bibitem[{\citenamefont{Yoshida}(2013{\natexlab{a}})}]{Beni11b}
\bibinfo{author}{\bibfnamefont{B.}~\bibnamefont{Yoshida}},
  \bibinfo{journal}{Ann. Phys. (NY)} \textbf{\bibinfo{volume}{338}},
  \bibinfo{pages}{134} (\bibinfo{year}{2013}{\natexlab{a}}).

\bibitem[{\citenamefont{Haah}(2011)}]{Haah11}
\bibinfo{author}{\bibfnamefont{J.}~\bibnamefont{Haah}}, \bibinfo{journal}{Phys.
  Rev. A} \textbf{\bibinfo{volume}{83}}, \bibinfo{pages}{042330}
  (\bibinfo{year}{2011}).

\bibitem[{\citenamefont{Yoshida}(2013{\natexlab{b}})}]{Beni13}
\bibinfo{author}{\bibfnamefont{B.}~\bibnamefont{Yoshida}},
  \bibinfo{journal}{Phys. Rev. B} \textbf{\bibinfo{volume}{88}},
  \bibinfo{pages}{125122} (\bibinfo{year}{2013}{\natexlab{b}}).

\bibitem[{\citenamefont{Yoshida}(2011)}]{Beni10b}
\bibinfo{author}{\bibfnamefont{B.}~\bibnamefont{Yoshida}},
  \bibinfo{journal}{Ann. Phys.} \textbf{\bibinfo{volume}{326}},
  \bibinfo{pages}{15} (\bibinfo{year}{2011}).

\bibitem[{\citenamefont{Fernandez-Pacheco}(1979)}]{Fernandez-Pacheco79}
\bibinfo{author}{\bibfnamefont{A.}~\bibnamefont{Fernandez-Pacheco}},
  \bibinfo{journal}{Phys. Rev. D} \textbf{\bibinfo{volume}{19}},
  \bibinfo{pages}{3173} (\bibinfo{year}{1979}).

\bibitem[{\citenamefont{Kubica and Yoshida}()}]{Kubica14}
\bibinfo{author}{\bibfnamefont{A.}~\bibnamefont{Kubica}} \bibnamefont{and}
  \bibinfo{author}{\bibfnamefont{B.}~\bibnamefont{Yoshida}},
  \eprint{arXiv:1402.0619}.

\bibitem[{\citenamefont{Albert and Barab{\'a}si}(2002)}]{Albert02}
\bibinfo{author}{\bibfnamefont{R.}~\bibnamefont{Albert}} \bibnamefont{and}
  \bibinfo{author}{\bibfnamefont{A.-L.} \bibnamefont{Barab{\'a}si}},
  \bibinfo{journal}{Rev. Mod. Phys.} \textbf{\bibinfo{volume}{74}},
  \bibinfo{pages}{47} (\bibinfo{year}{2002}).

\bibitem[{\citenamefont{Hamer}(2000)}]{Hamer00}
\bibinfo{author}{\bibfnamefont{C.~J.} \bibnamefont{Hamer}},
  \bibinfo{journal}{J. Phys. A: Math. Gen.} \textbf{\bibinfo{volume}{33}},
  \bibinfo{pages}{6683} (\bibinfo{year}{2000}).

\bibitem[{\citenamefont{Guo et~al.}(1994)\citenamefont{Guo, Bhatt, and
  Huse}}]{Guo94}
\bibinfo{author}{\bibfnamefont{M.}~\bibnamefont{Guo}},
  \bibinfo{author}{\bibfnamefont{R.~N.} \bibnamefont{Bhatt}}, \bibnamefont{and}
  \bibinfo{author}{\bibfnamefont{D.~A.} \bibnamefont{Huse}},
  \bibinfo{journal}{Phys. Rev. Lett.} \textbf{\bibinfo{volume}{72}},
  \bibinfo{pages}{4137} (\bibinfo{year}{1994}).

\end{thebibliography}
\end{document}